\documentclass{INTERSPEECH2025}

\usepackage{multirow}

\makeatletter
\newcommand*\concat
  {\mathbin{\mathmakebox[\widthof{${+}\m@th$}]{+\hskip-1emplus1fil+}}}

\makeatother

 \newcommand\citep{\cite}

\usepackage{soulutf8}
\usepackage{todonotes}
\usepackage{placeins}
\usepackage{tikz}
\usepackage{svg}

\interspeechcameraready

\title{{NanoCodec: Towards High-Quality Ultra Fast Speech LLM Inference}
\vspace{-0.08cm}
}

\author{Edresson Casanova$^{1}$, Paarth Neekhara$^{1}$, Ryan Langman$^{1}$, Shehzeen Hussain$^{1}$,\\ Subhankar Ghosh$^{1}$, Xuesong Yang$^{1}$,  Ante Jukić$^{1}$, Jason Li$^{1}$, Boris Ginsburg$^{1}$}

\affiliation{}{NVIDIA Corporation}{USA}

\address{
    {NVIDIA Corporation}
  }
  \email{\{ecasanova,pneekhara,rlangman,shehzeenh,subhankarg,xueyang,ajukic,jasoli,bginsburg\}@nvidia.com\vspace{-0.7cm}}


\begin{document}

\maketitle

\begin{abstract}
    {
Large Language Models (LLMs) have significantly advanced audio processing by leveraging audio codecs to discretize audio into tokens, enabling the application of language modeling techniques to speech data. However, existing audio codecs often operate at high frame rates, leading to slow training and inference, particularly for autoregressive models. To address this, there is growing interest in low frame-rate audio codecs, which reduce the number of autoregressive steps required to generate one second of audio. In this paper, we conduct ablation studies to examine the impact of frame rate, bitrate, and causality on codec reconstruction quality. Based on our findings, we introduce NanoCodec, a state-of-the-art audio codec that achieves high-quality compression at just 12.5 frames per second (FPS). NanoCodec outperforms related works across various bitrate ranges, establishing a new benchmark for low-latency and efficient Speech LLM training and inference.
    }
\end{abstract}

% \edresson{ For INTERSPEECH we need to have  5 pages  normally 4 pages for content and 1 for references}
% \vspace{-0.4cm}
\vspace{-0.12cm}
\section{Introduction}\label{sec:intro}

% \edresson{ I need to change the focus of introduction and abstract to remove the grouped code modeling stuff and make it more our codec centred.}
% Audio codec is an important signal processing technique, that compresses audio signals into discrete codes and then uses these codes to reconstruct the original audio. This technology has long held a central position in fields such as audio transmission and communication \cite{ai2024apcodec}. Recently, it also has been applied to some downstream tasks. For example, some researchers use the discrete codes generated by audio codecs combined with large language models (LLMs), to achieve impressive results in zero-shot text-to-speech (ZS-TTS) \cite{borsos2023audiolm, wang2023neural, zhang2023speechtokenizer, neekhara2024improving} and Speech-to-speech translation (S2ST) \cite{li2023textless, kim2023many, wei2023joint}. % inaguma2022unity

Audio coding is an important signal processing technique, that compresses audio signals into discrete codes and then uses these codes to reconstruct the original audio. These technologies have been central to fields such as audio transmission and communication \cite{ai2024apcodec, casanova2024low}. Recently, audio codecs have been applied to various downstream tasks. For instance, discrete codes generated by audio codecs combined with large language models (LLMs) have shown impressive results in tasks like zero-shot text-to-speech (ZS-TTS) \cite{zhang2023speechtokenizer, koel} and speech-to-speech translation (S2ST) \cite{li2023textless, kim2023many}.

In recent years, Neural Audio Codecs (NACs), which process raw waveforms as both input and output, have emerged as a promising solution that balances audio quality and bitrate \cite{zeghidour2021soundstream, defossez2022high, langman2024spectral, casanova2024low}. Notable examples include SoundStream \cite{zeghidour2021soundstream}, Encodec \cite{defossez2022high}, DAC \cite{kumar2024high}, and APCodec \cite{ai2024apcodec}. These models utilize Residual Vector Quantization (RVQ) \cite{vasuki2006review} to encode audio at low bitrates while preserving fidelity, often incorporating generative adversarial network (GAN)-based losses to enhance reconstruction quality. Research efforts have primarily focused on improving quantization strategies to enhance NAC performance. Several studies have refined RVQ techniques \cite{luo2024gull} and explored alternative approaches such as Finite Scalar Quantization (FSQ) \cite{mentzer2023finite, langman2024spectral, casanova2024low}. In addition, task-specific optimizations have been introduced, including integration of semantic information into quantization \cite{zhang2023speechtokenizer}. Further improvements have been achieved through architectural modifications \cite{xu2024intra} and reductions in computational complexity \cite{xu2024lightcodec}, collectively advancing the efficiency and effectiveness of NACs.
Additionally, researchers have pursued significant reductions in bitrate. For instance, WavTokenizer~\cite{ji2024wavtokenizer} encoded 24kHz audio at 0.9~kbps using a single codebook at 75~FPS, achieving audio quality comparable to previous NACs at a similar bitrate. TAAE~\cite{parker2024scaling} encodes 16~kHz audio at 25~FPS with bitrates of 0.4 and 0.7~kbps, surpassing previous works trained at the same bitrate.
TS3-Codec~\cite{wu2024ts3} encodes 16~kHz audio at 40-50~FPS with 0.68kbps-0.85kbps, achieving significantly better audio quality and computational efficiency than previous models at a similar bitrate.

Recently, there has been growing interest in reducing the frame rate of NACs, aiming to reduce the number of autoregressive steps required to generate audio in Speech LLMs. Notable examples include Mimi \cite{defossez2024moshi}, a model designed for duplex speech-to-speech applications, which compresses 24 kHz audio at 1.1 kbps and 12.5 FPS while maintaining reasonable audio quality. % However, while Mimi produces high-quality output for speakers seen during training, it often struggles to generalize to unseen speakers and multilingual data. 
The Low Frame-rate Speech Codec (LFSC) \cite{casanova2024low} provides high-quality 22.05 kHz audio compression at 1.89 kbps with a frame rate of 21.5 FPS. More recently, LFSC was integrated into the Koel-TTS model \cite{koel}, resulting in a threefold acceleration in inference for autoregressive zero-shot speech synthesis, while achieving state-of-the-art (SOTA) zero-shot performance in intelligibility, speaker similarity, and audio quality. This highlights the significant advantages of frame rate reduction.

% \edresson{Cite Low frame rate codec, Mimi, TS3-Codec, TAAE, MARS6 https://arxiv.org/abs/2501.05787}

Despite notable advancements in the field, several recent studies have overlooked the critical need for low latency, which presents significant challenges for enabling real-time streamable inference \cite{ai2024apcodec, xu2024intra, casanova2024low}. Achieving low latency in Speech LLMs can be accomplished through frame rate reduction \cite{casanova2024low}, architectural optimizations \cite{wu2024ts3}, and the incorporation of a causal decoder \cite{defossez2024moshi}, which facilitates inference without requiring token/frame caching. In this paper, we conduct a series of ablation studies to investigate the effects of frame rate, bitrate, and causality on codec reconstruction quality. Based on these insights, we propose NanoCodec, a SOTA audio codec that delivers high-quality compression at a remarkably low frame rate of 12.5 FPS, while outperforming related works across various bitrate ranges.

The key contributions of this work are as follows:
\begin{itemize}  
\vspace{-0.05cm}  
\item We introduce \textit{NanoCodec}, a novel state-of-the-art, partially causal audio codec that delivers high-quality audio compression at an exceptionally low frame rate of 12.5 FPS.
\item We conduct extensive ablation studies on frame rate reduction, causal audio coding, and the impact of bitrate on our proposed codec. 
% \item To demonstrate the effectiveness of our codec in delivering high-quality audio while reducing inference time for Speech LLMs, we train and evaluate a SOTA LLM-based TTS model, showcasing its practical benefits. 
\item To demonstrate our codec's effectiveness in enhancing audio quality and reducing inference time for Speech LLMs, we train and evaluate a SOTA LLM-based TTS model.
% \item Our codec implementation is publicly available in the NeMo repository\footnote{{https://github.com/NVIDIA/NeMo}}, and the pretrained checkpoints can be accessed on Hugging Face\footnote{{https://huggingface.co/nvidia/nanocodec-22khz-1.78kbps-12.5fps}}. % in the NeMo repository\footnote{\url{https://github.com/NVIDIA/NeMo}}.  
\item Our codec implementation is publicly available in the NeMo repository\footnote{{https://github.com/NVIDIA/NeMo}}, and the pretrained checkpoints can be accessed on Hugging Face\footnote{{https://huggingface.co/nvidia/nemo-nano-codec-22khz-1.78kbps-12.5fps}}.

\end{itemize}  

The audio samples for each of our experiments are available on the demo website\footnote{https://edresson.github.io/NanoCodec}.
% The audio samples for each of our experiments are available on the demo website\footnote{https://double-blind-review-demo.github.io/NanoCodec/}.

\section{NanoCodec}\label{sec:model}
NanoCodec builds upon the LFSC \cite{casanova2024low}, incorporating architectural modifications to enhance audio quality and intelligibility, while reducing the model number of parameters. The model consists of a fully convolutional generator network and three discriminators.  

The generator includes an encoder, vector quantization, and a HiFi-GAN-based decoder \cite{kong2020hifi}. The encoder features five residual blocks \cite{kong2020hifi}, each comprising three residual layers similar to the multi-receptive field fusion (MRF) module proposed by \cite{kong2020hifi}. Unlike \cite{casanova2024low}, which used a dilation rate of 1, we adopted dilation rates of 1, 3, and 5, as suggested by \cite{langman2024spectral}. We also explored various frame rates for the codec. To achieve a frame rate of 21.5, as in \cite{casanova2024low}, each residual block is followed by a 1D convolutional layer with strides of [2, 2, 4, 8, 8]. For frame rates of 25, 12.5, and 6.25, we used strides of [2, 3, 3, 7, 7], [2, 3, 6, 7, 7], and [3, 4, 6, 7, 7], respectively. The encoder has 24 initial channels, which are doubled after each downsampling 1D convolutional layer, and has a total of 30.4M parameters.

The decoder, based on the HiFi-GAN vocoder, uses upsampling rates that are the reverse of the encoder's  1D convolutional strides, following the approach in \cite{casanova2024low}. However, our decoder has 864 initial channels, which are halved after each upsampling layer. Inspired by \cite{lee2022bigvgan}, we replaced the Leaky ReLU activation in the decoder with Snake activation \cite{ziyin2020neural}. The decoder contains 31.6M parameters.

% The generator includes an encoder, vector quantization, and a HiFi-GAN-based decoder \cite{kong2020hifi}. The encoder features five residual blocks \cite{kong2020hifi}, each comprising three residual layers similar to the multi-receptive field fusion (MRF) module proposed by \cite{kong2020hifi}. Unlike \cite{casanova2024low}, which used a dilation rate of 1, we adopted dilation rates of 1, 3, and 5, as suggested by \cite{langman2024spectral}. We also explored various frame rates for the codec. To achieve a frame rate of 21.5, as in \cite{casanova2024low}, each residual block is followed by a 1D convolutional layer with strides of [2, 2, 4, 8, 8]. For frame rates of 25, 12.5, and 6.25, we used strides of [2, 3, 3, 7, 7], [2, 3, 6, 7, 7], and [3, 4, 6, 7, 7], respectively. The encoder contains 30.4M parameters.

% The decoder, based on the HiFi-GAN vocoder, uses upsampling rates that are the reverse of the encoder's  1D convolutional strides, following the approach in \cite{casanova2024low}. However, our encoder has 24 initial channels, which are doubled after each downsampling layer, while the decoder has 864 initial channels, which are halved after each upsampling layer. Inspired by \cite{lee2022bigvgan}, we replaced the Leaky ReLU activation in the decoder with Snake activation \cite{ziyin2020neural}. The decoder contains 31.6M parameters.

Additionally, we explored a causal variant of the encoder and decoder by replacing noncausal convolutional layers with causal ones, as in \cite{defossez2024moshi}. In total, these changes reduce the generator's parameter count by 45\% compared to LFSC.
%The encoder and decoder contain approximately 30.4M and 31.6M parameters, respectively, reducing the total parameter count by 45\% compared to LFSC. 

For the discriminators, we employ three neural networks using squared-GAN and feature-matching loss, consistent with \cite{casanova2024low}. We retained the WavLM-based discriminator and the multi-period discriminator \cite{kong2020hifi} from the original work \cite{casanova2024low}, but replaced the multi-scale complex STFT discriminator \cite{defossez2022high} with the multi-band multi-scale STFT discriminator \cite{kumar2024high}. 

In addition, inspired by \cite{yourtts}, we incorporate Speaker Consistency Loss (SCL) to improve speaker similarity in reconstructed speech. Specifically, we employ a pre-trained speaker encoder \cite{heo2020clova} to extract speaker embeddings from both the generated and ground truth audio, maximizing their cosine similarity.  Formally, let $\phi(\cdot)$ denote the function that outputs a speaker embedding, $cos\_sim$ the cosine similarity function, $\alpha$ a positive real-valued scaling factor controlling the contribution of SCL to the overall loss, and $n$ the batch size. The SCL is then defined as:

\vspace{-0.4cm}
\begin{equation}
\label{eq:SCL}
L_{\text{SCL}} =  \frac{- \alpha }{n} \sum_{i=1}^{n} \operatorname{cos\_sim}(\phi(g_{i}), \phi(h_{i})),
\end{equation}
\vspace{-0.1cm}
where $g_{i}$ and $h_{i}$ correspond to the ground truth and generated audio, respectively. In our experiments, we set $\alpha = 0.1$.

For vector quantization, we followed \cite{casanova2024low}, using Finite Scalar Quantization (FSQ) with eight codebooks, each containing four dimensions. The codebook levels are set to [8, 7, 6, 6], resulting in a total of 2016 codes per codebook. % Additionally, we explored different codebook levels to meet the specific requirements of our ablation studies.
Additionally, we investigated various configurations of codebook numbers and levels to meet the specific requirements of our ablation studies.

\subsection{Datasets}\label{sec:train-datasets}

% For our codec training, we utilized the same datasets employed in \cite{langman2024spectral}. This includes a 22.05kHz bandwidth-filtered version of the Common Voice 13 \cite{ardila2020common} training set and English audiobook-like data from the MLS dataset \cite{pratap2020mls}. The Common Voice-based training set comprises 105 languages, totaling 2.7 million utterances, and 3.2k hours of audio from 100 thousand speakers. The MLS English training dataset consists of  6.2 million utterances and 25.5k hours of audio from 4329 speakers.

For codec training, we utilize the two 22.05kHz audio datasets used in \cite{casanova2024low}. The Common Voice derived \cite{ardila2020common} training set comprises 105 languages, totaling 2.7 million utterances, and 3.2k hours of audio from about one-hundred thousand speakers. The MLS English training dataset consists of 6.2 million utterances and 25.5k hours of audio from 4329 speakers.

\subsection{Training setup}\label{sec:train}

We trained our codec for approximately 196,000 steps using 48 A100 GPUs with a batch size of 32. The total accumulated batch size was 1,536, and the model processed roughly 301 million samples. The training was conducted on 1.1-second audio excerpts. We used the Adam optimizer \cite{kingma2014adam} for both the generator and the discriminator, with $\beta_1 = 0.8$, $\beta_2 = 0.99$, and an initial learning rate of 2e-4, which decayed exponentially with a gamma of 0.998.

\subsection{Results and Discussion}\label{sec:results}

% We followed an evaluation strategy similar to \cite{langman2024spectral} and \cite{casanova2024low}. For evaluating perceptual quality, we estimate Mean Opinion Scores (MOS) using Torchaudio-Squim \cite{kumar2023torchaudio}. Time-domain accuracy is measured using SI-SDR \cite{le2019sdr}. Spectral accuracy is assessed by calculating the L1 distance between log mel-spectrogram (Mel Dist.) feature. To measure the intelligibility of the codecs reconstruction we compute the character error rate (CER) between the \mbox{Massively Multilingual Speech (MMS) model} \cite{pratap2024scaling} transcriptions of the ground truth audio and the reconstructed audio. Speaker similarity was assessed by calculating the Speaker Encoder Cosine Similarity (SECS) \cite{yourtts} using the SOTA ECAPA2 \cite{thienpondt2024ecapa2} speaker encoder.

We followed an evaluation strategy similar to \cite{casanova2024low}. To assess perceptual quality, we estimate Mean Opinion Scores (MOS) using Torchaudio-Squim \cite{kumar2023torchaudio} (SQMOS) and the wide-band Perceptual Evaluation of Speech Quality (PESQ) metric \cite{rix2001perceptual}. Spectral accuracy is evaluated by computing the L1 distance between log mel-spectrograms (Mel Dist.). To quantify the intelligibility of the codec's reconstruction, we compute the Character Error Rate (CER) by comparing transcriptions from the Massively Multilingual Speech (MMS) model \cite{pratap2024scaling} for both the ground-truth and reconstructed audio. Speaker similarity is assessed using the Speaker Encoder Cosine Similarity (SECS) \cite{yourtts}, computed with the SOTA ECAPA2 speaker encoder \cite{thienpondt2024ecapa2}.

For evaluation, we use the 44.1kHz MLS test set \cite{casanova2024low}, which includes 200 samples per language across eight languages. Additionally, we assess performance on the F10 and M10 speakers from the DAPS clean dataset to evaluate model quality on studio-recorded audio.

% For evaluation, we use the 44.1kHz MLS dataset test set \cite{pratap2020mls}, as released by \cite{casanova2024low}. This dataset comprises 200 samples from each of the eight MLS languages, making it a suitable choice given that most contemporary Speech LLM models are trained on audiobook-style data. Additionally, we evaluate the models on the F10 and M10 speakers from the DAPS clean dataset, which was previously used for assessing the DAC and LFSC models. This inclusion allows us to examine model performance on studio-quality audio.  

\subsubsection{Ablation Study}\label{sec:ablations}

We conduct an ablation study to analyze the enhancements introduced to the LFSC architecture, as well as the effects of causality, frame rate, and bitrate reduction on our codec. Table \ref{tab:codec-results-full} presents the evaluation results on the 44.1kHz MLS test set and the F10 and M10 DAPS speakers. For consistency, both ground-truth and reconstructed audio were downsampled to 16kHz during metric computation. % to ensure a fair comparison against related works.

\begin{table*}% [!ht]
\vspace{-0.2cm}
\caption{Reconstruction performance of different codec configurations using MLS 44.1kHz test set and F10 and M10 DAPS speakers}
\vspace{-0.2cm}
\label{tab:codec-results-full}
\centering
\resizebox{0.999\textwidth}{!}{%
\begin{tabular}{l|c|c|c|c|c|c|c|c|c|c|c|c|c|c}
\toprule
\textbf{Codec}                                                                                     & \textbf{Sampling rate}    & \textbf{Bitrate}                   & \textbf{Token/Sec}            & \textbf{Frames/Sec}   & \textbf{\# Codebooks}  & \textbf{\# Codes}  &  \textbf{Encoder} & \textbf{Decoder} &  \textbf{Dataset} & \textbf{SQMOS ($\uparrow$)} & \textbf{PESQ($\uparrow$)} & \textbf{Mel Dist. ($\downarrow$)} &   \textbf{SECS} ($\uparrow$)& \textbf{CER($\downarrow$)}  \\ \hline

\multirow{2}{*}{\begin{tabular}[c]{@{}l@{}} LFSC\cite{casanova2024low}\end{tabular}}                                                                              & \multirow{2}{*}{22.05kHz} & \multirow{2}{*}{{1.89kbps}} & \multirow{2}{*}{{172}} & \multirow{2}{*}{{21.5}}& \multirow{2}{*}{8} &  \multirow{2}{*}{2016} & \multirow{2}{*}{Noncausal} & \multirow{2}{*}{Noncausal} & MLS              &\textbf{4.432} & 2.831& 0.146 & 0.841 & 2.529\\      
                                                                                                   &                           &                                    &                               &                &  &                & && DAPS      &      4.677& 3.122 &  0.141 & 0.807 & {0.655}          \\ \hline

\multirow{2}{*}{Ours}                                                                              & \multirow{2}{*}{22.05kHz} & \multirow{2}{*}{{1.89kbps}} & \multirow{2}{*}{{172}} & \multirow{2}{*}{{21.5}}& \multirow{2}{*}{8} &  \multirow{2}{*}{2016} &  \multirow{2}{*}{Noncausal} & \multirow{2}{*}{Noncausal} & MLS              &      4.427    &         \textbf{ 2.919}      &            \textbf{0.136}           &                                   \textbf{0.876}   &     \textbf{2.203}             \\
                                                                                                   &                           &                                    &                  & &             &                                 & && DAPS             & \textbf{{4.693}  }                  &      \textbf{3.271}                   & \textbf{{0.130} }                                       &       \textbf{0.844 }     &       \textbf{{0.577}}              \\ \hline\hline

\multirow{2}{*}{Ours}                                                                              & \multirow{2}{*}{22.05kHz} & \multirow{2}{*}{{1.1kbps}} & \multirow{2}{*}{{100}} & \multirow{2}{*}{{12.5}} & \multirow{2}{*}{8}& \multirow{2}{*}{2016} &\multirow{2}{*}{Causal} & \multirow{2}{*}{Causal} & MLS              &          4.394                  &    1.926                   & 0.211                            &  0.735   &    5.490 \\
                   &                           &                                    &        &                        &            &                     & && DAPS             &                     4.655                        &     2.058                     &           0.2157                 &         0.717      &        1.124\\ \hline
                   
\multirow{2}{*}{Ours}                                                                              & \multirow{2}{*}{22.05kHz} & \multirow{2}{*}{{1.1kbps}} & \multirow{2}{*}{{100}} & \multirow{2}{*}{{12.5}} & \multirow{2}{*}{8}& \multirow{2}{*}{2016} & \multirow{2}{*}{Noncausal} & \multirow{2}{*}{Noncausal} & MLS              & {4.423}                             &          \textbf{2.592 }             & \textbf{{0.158}}                            & \textbf{{0.825}}                             &                \textbf{3.614}     \\
                   &                           &                                    &       & &                                                &       & && DAPS             &        \textbf{ 4.683 }           &   \textbf{2.893}                                                &             \textbf{0.154 }              &          \textbf{0.785}     &      {0.927} \\ \hline

\multirow{2}{*}{Ours}                                                                              & \multirow{2}{*}{22.05kHz} & \multirow{2}{*}{{1.1kbps}} & \multirow{2}{*}{{100}} & \multirow{2}{*}{{12.5}}& \multirow{2}{*}{8} &  \multirow{2}{*}{2016} & \multirow{2}{*}{Noncausal} & \multirow{2}{*}{Causal} & MLS              &    \textbf{4.426} & {2.378} & {0.177} & {0.788}& {3.617}         \\
                   &                           &                                    &       & &                                        &               & && DAPS            &{ 4.564}&   {2.646} & {0.171} & {0.777} & {0.839} \\ \hline \hline

% \multirow{2}{*}{Ours}                                                                              & \multirow{2}{*}{22.05kHz} & \multirow{2}{*}{{1.1kbps}} & \multirow{2}{*}{{100}} & \multirow{2}{*}{{12.5}}& \multirow{2}{*}{8} &  \multirow{2}{*}{2016} & \multirow{2}{*}{Causal} & \multirow{2}{*}{Noncausal} & MLS              & {4.220}                             & {-9.643}                        & {0.248}                            & {{0.654}}                            &                {8.604}      \\
%                    &                           &                                    &       & &                                        &               & && DAPS             &     4.566                &              -7.753           &                        0.253    &                0.656            &                 2.492     \\ \hline
                   
                   \hline

% \multirow{2}{*}{Ours}                                                                              & \multirow{2}{*}{22.05kHz} & \multirow{2}{*}{{1.1kbps}} & \multirow{2}{*}{{100.22}} & \multirow{2}{*}{{100.22}} & \multirow{2}{*}{1} & \multirow{2}{*}{Noncausal} & \multirow{2}{*}{Causal} & MLS              &         4.414                   &          1.786             &             0.188              &                       0.724     &                  6.289\\
%                    &                           &                                    &       & &                        &                    &            && DAPS             &        4.675             &                                       4.458            &              0.184              &      0.716         &    1.201    \\ \hline
                   
% \multirow{2}{*}{Ours}                                                                              & \multirow{2}{*}{22.05kHz} & \multirow{2}{*}{{1.1kbps}} & \multirow{2}{*}{{100}} & \multirow{2}{*}{{50}} & \multirow{2}{*}{2} & \multirow{2}{*}{Noncausal} & \multirow{2}{*}{Causal} & MLS              &             4.418               &      0.718                 &           0.178                &                       0.768      &                  4.661\\
%                    &                           &                                    &       & &                        &                    &            && DAPS             &         4.641            &        3.421                                           &                0.174            &         0.738      &      1.038  \\ \hline

\multirow{2}{*}{Ours}                                                                              & \multirow{2}{*}{22.05kHz} & \multirow{2}{*}{{1.1kbps}} & \multirow{2}{*}{{100}} & \multirow{2}{*}{{25}}  & \multirow{2}{*}{4}& \multirow{2}{*}{2016} & \multirow{2}{*}{Noncausal} & \multirow{2}{*}{Causal}  & MLS              &             4.415               &            \textbf{2.534}          &          \textbf{0.167}                &   \textbf{0.798}                          &               {3.749}   \\
                   &                           &                                    &            & &                                      &            & && DAPS             &      \textbf{4.662 }              &    \textbf{2.863  }                                         &    \textbf{0.160 }                          &     {0.755}          &      \textbf{0.622}  \\ \hline 

\multirow{2}{*}{Ours}                                                                              & \multirow{2}{*}{22.05kHz} & \multirow{2}{*}{{1.1kbps}} & \multirow{2}{*}{{100}} & \multirow{2}{*}{{12.5}}& \multirow{2}{*}{8} &  \multirow{2}{*}{2016} & \multirow{2}{*}{Noncausal} & \multirow{2}{*}{Causal} & MLS              &    \textbf{{4.426}} & {2.378} & {0.177} & {0.788}& \textbf{{3.617}}       \\
                   &                           &                                    &       & &                                        &               & && DAPS             &         { 4.564}&   {2.646} & {0.171} & \textbf{{0.777}} & {0.839}  \\ \hline

\multirow{2}{*}{Ours}                                                                              & \multirow{2}{*}{22.05kHz} & \multirow{2}{*}{{1.1kbps}} & \multirow{2}{*}{{100}} & \multirow{2}{*}{{6.25}}  & \multirow{2}{*}{16}& \multirow{2}{*}{2016} &\multirow{2}{*}{Noncausal} & \multirow{2}{*}{Causal} & MLS              & {4.400}                             &              2.081          & 0.200                           & 0.746                            &                {5.293}      \\
                   &                           &                                    &               & &                  &                      &          && DAPS             &                       4.652                      &              2.135             &                0.213            &      0.711         &    1.419    \\ \hline\hline

\multirow{2}{*}{Ours}                                                                              & \multirow{2}{*}{22.05kHz} & \multirow{2}{*}{{1.78kbps}} & \multirow{2}{*}{{162.5}} & \multirow{2}{*}{{12.5}}& \multirow{2}{*}{13} & \multirow{2}{*}{2016} &\multirow{2}{*}{Noncausal} & \multirow{2}{*}{Causal} & MLS              &            \textbf{4.441}  & \textbf{2.760}& \textbf{0.143} &  \textbf{0.862} &  \textbf{2.423}        \\
                   &                           &                                    &       & &                                        &               & && DAPS             &    \textbf{4.697} & \textbf{3.030} &\textbf{ 0.139} & \textbf{0.831} & \textbf{0.758}     \\ \hline

\multirow{2}{*}{Ours}                                                                              & \multirow{2}{*}{22.05kHz} & \multirow{2}{*}{{1.1kbps}} & \multirow{2}{*}{{100}} & \multirow{2}{*}{{12.5}}& \multirow{2}{*}{8} &  \multirow{2}{*}{2016} & \multirow{2}{*}{Noncausal} & \multirow{2}{*}{Causal} & MLS              &    {{4.426}} & {2.378} & {0.177} & {0.788}& {{3.617}}       \\
                   &                           &                                    &       & &                                        &               & && DAPS             &         { 4.564}&   {2.646} & {0.171} & {0.777} & {0.839}  \\ \hline

\multirow{2}{*}{Ours}                                                                              & \multirow{2}{*}{22.05kHz} & \multirow{2}{*}{{0.8kbps}} & \multirow{2}{*}{{50}} & \multirow{2}{*}{{12.5}}& \multirow{2}{*}{4} & \multirow{2}{*}{65536} & \multirow{2}{*}{Noncausal} & \multirow{2}{*}{Causal} & MLS              &    4.369 &  {2.033} & {0.205} & {0.706} &  {6.519}      \\
                   &                           &                                    &       & &                                        &               & && DAPS             &        4.617 & {2.237} & {0.202} &  {0.645} & {1.189}  \\ \hline

\multirow{2}{*}{Ours}                                                                              & \multirow{2}{*}{22.05kHz} & \multirow{2}{*}{{0.6kbps}} & \multirow{2}{*}{{50}} & \multirow{2}{*}{{12.5}}& \multirow{2}{*}{4} & \multirow{2}{*}{4032} & \multirow{2}{*}{Noncausal} & \multirow{2}{*}{Causal} & MLS              &    {4.407} & 2.012 &  {0.205} & { 0.701} & {7.792} \\
       &                           &                                    &       & &                                        &               & && DAPS             &        {4.662} &  {2.205} &  {0.204} & {0.656} & {1.469}  \\ \hline\hline

\multirow{2}{*}{Ours}                                                                              & \multirow{2}{*}{22.05kHz} & \multirow{2}{*}{{1.78kbps}} & \multirow{2}{*}{{162.5}} & \multirow{2}{*}{{12.5}}& \multirow{2}{*}{13} & \multirow{2}{*}{2016} &\multirow{2}{*}{Noncausal} & \multirow{2}{*}{Causal} & MLS              &            \textbf{4.441}  & {2.760}& \textbf{0.143} &  \textbf{0.862} &  \textbf{2.423}        \\
                   &                           &                                    &       & &                                        &               & && DAPS             &    \textbf{4.697} & {3.030} &\textbf{ 0.139} & {0.831} & {0.758}     \\ \hline
\multirow{2}{*}{\begin{tabular}[c]{@{}l@{}} Ours\end{tabular}}                                                                              & \multirow{2}{*}{22.05kHz} & \multirow{2}{*}{{1.89kbps}} & \multirow{2}{*}{{172}} & \multirow{2}{*}{{21.5}}& \multirow{2}{*}{8} &  \multirow{2}{*}{2016} & \multirow{2}{*}{Noncausal} & \multirow{2}{*}{Causal} & MLS              & 4.427 & \textbf{2.837} &  0.150&  0.854& 2.434\\      
                                                                                                   &                           &                                    &                               &                &  &                & && DAPS      &  4.666    &\textbf{3.156} &  0.145 & \textbf{0.832} &        \textbf{0.601}   \\ \bottomrule

\end{tabular}
}
\vspace{-0.4cm}
\end{table*}

\textbf{Architectural Improvements:} The proposed architectural modifications enhanced the LFSC model across nearly all metrics while reducing the number of parameters by approximately 45\%.

\textbf{Causality:} 
A causal codec is preferable for Speech LLM models as it minimizes first-token latency. In contrast, the noncausal LFSC requires a lookahead of approximately five frames (232.5 ms), meaning that tokens from the first five autoregressive steps must be cached to achieve the same quality in both streaming and offline inference. This constraint introduces a significant delay in generating the initial audio output. For a demonstration of the impact of this lookahead requirement, please refer to our demo website\footnote{https://edresson.github.io/NanoCodec}. %\footnote{https://double-blind-review-demo.github.io/NanoCodec/}.
To address this issue, we explored different combinations of causal and non-causal encoder/decoder configurations. Following the Mimi codec \cite{defossez2024moshi}, we reduced the bitrate to 1.1 kbps and the frame rate to 12.5 FPS. Our fully causal model yielded the worst results, particularly in intelligibility and speaker similarity. Conversely, the fully non-causal model achieved the best performance across nearly all metrics. The partially causal model, which employs a non-causal encoder and a causal decoder, demonstrated competitive performance compared to the fully non-causal model. For certain Speech LLM tasks, such as TTS and Speech-to-Speech models where ASR or SSL models serve as the encoder, a causal encoder is unnecessary since only the codec decoder is used during inference. In these scenarios, our experiments suggest that a partially causal codec may be beneficial. Given that our primary goal in this paper is TTS downstream task, we will adopt the partially causal model for future experiments.

\textbf{Frame Rate Reduction:} % Lowering the frame rate is a desirable feature as it directly reduces the Speech LLM number of autoregressive steps required to generate one second of audio, thereby improving inference speed and reducing latency. To investigate its impact, we conducted three experiments, training our 1.1 kbps codec at 25, 12.5, and 6.25 FPS. Our results indicate that the 25 FPS codec outperforms the 12.5 FPS variant in PESQ and Mel Distance metrics. However, for overall perceptual quality (SQMOS), speaker similarity (SECS), and intelligibility (CER), no clear winner emerges between the two. 
% The 6.25 FPS codec, however, performs significantly worse, particularly in terms of intelligibility. We hypothesize that at 6.25 FPS, the model frequently compresses two distinct phonemes within a single frame, whereas at 12.5 FPS, such occurrences are considerably less frequent. Notably, most English speakers produce speech at a rate of approximately ten to twelve phonemes per second \cite{roach2002little}, which further supports this observation.
Reducing the frame rate in Speech LLMs decreases the number of autoregressive steps per second of audio, improving inference speed and reducing latency. To evaluate this effect, we trained our 1.1 kbps codec at 25, 12.5, and 6.25 FPS. While the 25 FPS codec outperformed the 12.5 FPS variant in PESQ and Mel Distance, both exhibited comparable performance in perceptual quality (SQMOS), speaker similarity (SECS), and intelligibility (CER). In contrast, the 6.25 FPS codec showed a significant decline, particularly in intelligibility. We hypothesize that at 6.25 FPS, the model frequently compresses two distinct phonemes within a single frame, whereas such occurrences are considerably less frequent at 12.5 FPS. Given that most English speakers produce speech at a rate of approximately ten to twelve phonemes per second \cite{roach2002little}, this observation is further supported.

\textbf{Bitrate Reduction:} Recent research has extensively explored bitrate reduction in neural audio codecs. The primary hypothesis is that lower bitrates yield simpler representations, making it easier for Speech LLMs to learn the target distribution. However, there is an inherent trade-off between bitrate, audio quality, and intelligibility. To further investigate this trade-off, we trained our 12.5 FPS codec at 1.78 kbps, 1.1 kbps, 0.8 kbps, and 0.6 kbps. As expected, higher-bitrate models produced better results across all metrics. Notably, bitrate reduction has a substantial impact on intelligibility (CER), whereas general audio quality and speaker similarity are comparatively less affected. Specifically, reducing the bitrate from 1.78 kbps to 1.1 kbps led to a relative degradation of approximately 1.49 times in CER (from 2.423 to 3.617). A further reduction to 0.8 kbps resulted in a 2.69 times relative increase in CER (from 2.423 to 6.519). At 0.6 kbps, intelligibility deteriorated even further, with pronunciation errors becoming significantly more frequent. These findings strongly suggest that such a decline in intelligibility could severely impact the performance of Speech LLM models trained with extremely low-bitrate codecs. 

\textbf{21.5 FPS vs. 12.5 FPS:} We further compared our model trained at 1.78 kbps with a frame rate of 12.5 FPS against the same model trained at 1.89 kbps with a frame rate of 21.5 FPS. Both models show comparable performance, with 12.5 FPS excelling in SQMOS and 21.5 FPS in PESQ. % Speaker similarity and intelligibility remain almost unchanged, indicating that lower frame rates can maintain quality while improving efficiency.
% For speaker similarity and intelligibility, there is no clear winner, suggesting that lower frame rates can preserve quality while enhancing efficiency.
Speaker similarity and intelligibility remain comparable, indicating that lower frame rates can maintain quality while improving efficiency.

%  \textbf{21.5 FPS vs. 12.5 FPS:} Models trained at 1.78 kbps (12.5 FPS) and 1.89 kbps (21.5 FPS) demonstrate comparable performance, with 12.5 FPS achieving higher SQMOS and 21.5 FPS excelling in PESQ, while speaker similarity and intelligibility remain unchanged, suggesting that lower frame rates can enhance efficiency without compromising quality.

% The results align with our previous analysis at 1.1 kbps, showing that both models achieve comparable performance. While the 12.5 FPS model performs better in SQMOS, the 21.5 FPS model achieves higher PESQ scores. For speaker similarity and intelligibility, no clear winner emerges, indicating that reducing the frame rate can maintain competitive performance while improving computational efficiency.

% # Bit rates:
% Spectral codec:  10 bits that store 1024 values  possible values (we have 1000) * 688 =  ~6.88kbps
% Low Frame rate Speech Codec:  11 bits that store 2048 possible values values (we have 2016) * 172 (21.5 * 8)  =  ~1.85kbps
% DAC 44khz: 10 bits that store 1024 values  possible values * 688 =  ~7.75kbps
% DAC 22khz: 10 bits that store 1024 values  possible values * 2400 =  24kbps
% APCodec + APBWE: 10 bits that store 1024 values  possible values * x =  1kbps
% Mini 8 codebooks: 11 * 100 = 1.1kbps
% Mini 32 codebooks: 11 * 400 = 1.1kbps

% NeMo Speech Codec:  11 bits that store 2048 possible values values (we have 2016) * 172 (21.5 * 8)  =  ~1.85kbps

\subsubsection{Comparison with Related Works}\label{sec:related-works}
To ensure a fair comparison with related works, we trained our model at a bitrate comparable to previous approaches. For LFSC~\cite{casanova2024low}, Mimi~\cite{defossez2024moshi}, WavTokenizer~\cite{ji2024wavtokenizer}, and TAAE~\cite{parker2024scaling}, we utilized publicly available checkpoints and inference code. For TS3-Codec~\cite{wu2024ts3}, we evaluated samples generously provided by the authors. For consistency, we downsampled both ground-truth and reconstructed audio to 16kHz during metric computation, ensuring a fair comparison with TAAE and TS3-Codec, which operate at this sampling rate. Table \ref{tab:related-works} presents the results of our evaluation on the 44.1~kHz MLS test set and the F10 and M10 DAPS speakers, highlighting the performance of our approach relative to state-of-the-art codecs.

\begin{table}% [!ht]
\vspace{-0.25cm}
% \caption{Reconstruction performance of different SOTA codecs using MLS 44.1kHz test set and F10 and M10 DAPS speakers}
\caption{Reconstruction performance of different SOTA codecs.}
\vspace{-0.25cm}
\label{tab:related-works}
\centering
\resizebox{0.47\textwidth}{!}{%
% \begin{tabular}{l|l|l|l|l|l|l|l|l|l|l}
\begin{tabular}{l|c|c|c|c|c|c|c}
\toprule
\textbf{Codec} & \textbf{Bitrate} &      \textbf{Dataset} & \textbf{SQMOS} & \textbf{PESQ} & \textbf{Mel Dist.} &   \textbf{SECS} & \textbf{CER}  \\ \hline

% \multirow{2}{*}{\begin{tabular}[c]{@{}l@{}} DAC\end{tabular}}& \multirow{2}{*}{7.75kbps} 
% & MLS & 4.432 & - & 0.136 & 0.838 & 2.09\\
% & &DAPS &  4.687& -&  0.129 & 0.810 & \textbf{0.55}\\ \hline

\multirow{2}{*}{\begin{tabular}[c]{@{}l@{}} LFSC \cite{casanova2024low}\end{tabular}}& \multirow{2}{*}{1.89kbps} 
& MLS & 4.432 & \textbf{2.831}& 0.146 & 0.841& 2.529\\
& &DAPS &  4.677& \textbf{3.122} &  0.141 & 0.807 & \textbf{0.655}\\ \hline

\multirow{2}{*}{\begin{tabular}[c]{@{}l@{}}Ours\end{tabular}}& \multirow{2}{*}{1.78kbps} 
& MLS & \textbf{4.441}  & 2.760& \textbf{0.143} &  \textbf{0.862} &  \textbf{2.423}\\
& &DAPS & \textbf{4.697} & 3.030 &\textbf{ 0.139} & \textbf{0.831} & 0.758\\ \hline

\hline

\multirow{2}{*}{\begin{tabular}[c]{@{}l@{}} Mimi \cite{defossez2024moshi} \end{tabular}}& \multirow{2}{*}{1.1kbps} 
& MLS & 4.333 & 2.132 & 0.238 & 0.678 & 7.221\\
& &DAPS & 4.302& 2.316 & 0.244 & 0.569 &1.629 \\ \hline
\multirow{2}{*}{\begin{tabular}[c]{@{}l@{}} Ours \end{tabular}}& \multirow{2}{*}{1.1kbps} 
& MLS &  \textbf{4.426} & \textbf{2.378} & \textbf{0.177} & \textbf{0.788}& \textbf{3.617}\\
& &DAPS &\textbf{ 4.564}&   \textbf{2.646} & \textbf{0.171} & \textbf{0.777} & \textbf{0.839}\\ \hline
\hline

\multirow{2}{*}{\begin{tabular}[c]{@{}l@{}}  WavTokenizer \cite{ji2024wavtokenizer} \end{tabular}}& \multirow{2}{*}{0.9kbps} 
& MLS & \textbf{4.404}  &  1.877 & 0.207 & 0.576& 15.27\\
& &DAPS & {4.671} &  2.222 & 0.213  & 0.535 & 4.042 \\ \hline
\multirow{2}{*}{\begin{tabular}[c]{@{}l@{}} TS3-Codec \cite{wu2024ts3} \end{tabular}}& \multirow{2}{*}{0.85kbps} 
& MLS & 4.392  &  \textbf{2.389}& \textbf{0.183} & 0.672 & 8.159\\
& &DAPS &  \textbf{4.674} & \textbf{2.770} &  \textbf{0.167} & \textbf{0.733} &  1.522\\ \hline
\multirow{2}{*}{\begin{tabular}[c]{@{}l@{}} Ours \end{tabular}}& \multirow{2}{*}{0.8kbps} 
& MLS &  4.369 &  {2.033} & {0.205} & \textbf{0.706} &  \textbf{6.519}\\
& &DAPS & 4.617 & {2.237} & {0.202} &  {0.645} & \textbf{1.189}\\ \hline
\hline

\multirow{2}{*}{\begin{tabular}[c]{@{}l@{}}  TAAE \cite{parker2024scaling} \end{tabular}}& \multirow{2}{*}{0.7kbps} 
& MLS & 4.370 & \textbf{2.053} &  0.320 &  0.323 & 18.355 \\
& &DAPS &  4.611& \textbf{2.250} & 0.248 & 0.372 & 3.692\\ \hline

\multirow{2}{*}{\begin{tabular}[c]{@{}l@{}} Ours \end{tabular}}& \multirow{2}{*}{0.6kbps} 
& MLS &  \textbf{4.407} & 2.012 &  \textbf{0.205} & \textbf{ 0.701} & \textbf{7.792} \\
& &DAPS & \textbf{4.662} &  {2.205} &  \textbf{0.204} & \textbf{0.656} & \textbf{1.469} \\ \bottomrule

\end{tabular}
}
\vspace{-0.68cm}
\end{table}

Our model, trained at 1.78 kbps, outperforms the 1.89 kbps LFSC across most metrics, while benefiting from the advantages of a causal decoder and approximately 45\% fewer parameters. At 1.1 kbps, our model outperforms the Mimi codec across all metrics, showing a particularly large improvement in intelligibility. At 0.8 kbps, while the overall audio quality of our model is slightly lower than that of the 0.9 kbps WavTokenizer, it consistently outperforms the WavTokenizer across all other evaluation metrics. When compared to the 0.85 kbps TS3-Codec, our 0.8 kbps model exhibits better intelligibility on both evaluation datasets. Additionally, it achieves superior speaker similarity on the multilingual MLS dataset, though it shows a slight reduction in speaker similarity for two high-quality English speakers in the DAPS dataset. Despite these minor trade-offs in specific metrics, our model delivers these results with a 50 bps smaller bitrate and is trained at 22.05 kHz, while the TS3-Codec was trained at 16 kHz. This distinction is significant, as our model encodes a broader frequency range. At 0.6 kbps, our model significantly outperforms the TAAE codec in most metrics, although TAAE slightly surpasses our model in PESQ. Overall, our model achieves SOTA results across a wide range of bitrates, demonstrating the efficiency and effectiveness of the proposed NanoCodec. Notably, it excels in enhancing key metrics such as intelligibility and speaker similarity, further solidifying its superiority in low-bitrate audio coding scenarios.

% \edresson{Evaluate and discuss 8k codes model}
% \section{TTS experiments}\label{sec:tts-exp}

\section{Zero-Shot TTS study}\label{sec:tts-exp}

\subsection{Experiments setup}\label{sec:tts-exps-setup}

% For training the TTS model, we employed the same datasets as those used in \cite{koel}. These datasets comprise 18k hours of English data from three sources: the train-clean 100 and 360 subsets of LibriTTS~\cite{zen2019libritts}, HiFiTTS~\cite{bakhturina21_interspeech}, and a 17k-hour subset of the LibriVox MLS dataset ~\cite{pratap2020mls}.%, and a proprietary dataset featuring two speakers and totaling 63 hours. 

To assess the performance of our codec in comparison to the LFSC and explore the effects of bitrate reduction, we adopted Koel-TTS~\cite{koel}, a SOTA LLM-based TTS model. % We employed the Koel-TTS model with context conditioning on the decoder and Classifier-Free Guidance (CFG) owing to its superior performance on the zero-shot TTS task. 
% We employed the Koel-TTS model with context conditioning on the decoder and Classifier-Free Guidance (CFG), due to its superior performance on the zero-shot TTS task.
% We employed the Decoder Context Koel-TTS variant with Classifier-Free Guidance (CFG), due to its superior performance on the zero-shot TTS task.
% We employed the Decoder Context Koel-TTS variant with Classifier-Free Guidance (CFG) for its superior performance in zero-shot TTS.
Among Koel-TTS architectures, we use the Decoder Context variant with Classifier-Free Guidance (CFG) as it achieves the best performance in zero-shot TTS.
For simplicity, we did not apply Direct Preference Optimization in our experiments. For training the TTS model, we used the same datasets as \cite{koel}. These datasets consist of 18,000 hours of English data from three sources: the train-clean 100 and 360 subsets of LibriTTS~\cite{zen2019libritts}, HiFiTTS~\cite{bakhturina21_interspeech}, and a 17,000-hour subset of the LibriVox MLS dataset~\cite{pratap2020mls}.

% The model was trained using both the 1.89 kbps LFSC and the proposed NanoCodec at bitrates of 1.89 kbps, 1.78 kbps, and 1.1 kbps. We employed a fixed context duration of five seconds, where the context consists of an alternate utterance from the same speaker as the target utterance. Each model was trained for approximately 200,000 steps using 32 A100 GPUs with a batch size of 6. The total accumulated batch size was 192. The training process used the AdamW optimizer with a fixed learning rate of 1e-4. During inference, we applied multinomial Top-k sampling with $K=80$,  CFG scale $2.5$, and a temperature of 0.6.

The model was trained using 1.89 kbps LFSC, as well as NanoCodec at 1.89 kbps, 1.78 kbps, and 1.1 kbps bitrates, using a fixed five-second context from an alternate utterance of the same speaker. Training ran for 200,000 steps on 32 A100 GPUs with a batch size of 8, totaling 256. The AdamW optimizer was used with a 1e-4 learning rate. During inference, we applied multinomial Top-k sampling with $K=80$,  CFG scale $2.5$, and a temperature of 0.6.

\subsection{Results and Discussion}\label{sec:tts-results}
%  \vspace{-0.35cm}

% To evaluate the performance of the Koel-TTS model with various codecs, we followed the methodology proposed by~\cite{koel}. We evaluated the model with unseen speakers, for that we have used the same subset of \textit{test-clean} LibriTTS containing $180$ utterances from a total of $36$ speakers. We use a random $5$ second slice from the context audio during inference for all experiments. CER was computed by aligning the transcription of the generated audio with the TTS input text using the Parakeet-TDT ASR model~\cite{xu2023efficient}. SECS was calculated using the Titanet-Small~\cite{koluguri2022titanet} speaker verification model.

To evaluate the performance of the Koel-TTS model with various codecs, we followed the methodology proposed by~\cite{koel}. We evaluated the model on unseen speakers, using the same subset of \textit{test-clean} LibriTTS containing $180$ utterances from a total of $36$ speakers. We use a random $5$ second slice from the context audio during inference for all experiments. CER was computed using the Parakeet-TDT ASR model~\cite{xu2023efficient}. SECS was calculated using the Titanet-Small~\cite{koluguri2022titanet} speaker verification model.

{For MOS evaluation, each sample was rated on a 5-point scale by at least 37 independent listeners.}
Additionally, we measured the inference real-time factor (RTF) and time-to-first-audio (TTFA) using a batch size of 16 on a single RTX 6000 GPU. To ensure consistency, both RTF and TTFA values were normalized by setting the fastest model’s values to 1 and scaling those of the other models accordingly.

\begin{table}[ht!]
\vspace{-0.25cm}
\caption{Koel-TTS evaluation using different codecs}
\vspace{-0.25cm}
\label{tab:results-tts}
\centering
\resizebox{0.49\textwidth}{!}{%
\begin{tabular}{l|l|l|l|l|l|l|l}
\toprule
 \textbf{Codec} &  \textbf{Bitrate} &  \textbf{Frames/Sec} & \textbf{MOS($\uparrow$)} & \textbf{CER($\downarrow$)} & \textbf{SECS($\uparrow$)} & \textbf{RTF($\downarrow$})   & \textbf{TTFA($\downarrow$}) \\ \hline
\begin{tabular}[c]{@{}l@{}} LFSC~\cite{koel}\end{tabular}   &   1.89kbps      &  21.5 &  \textbf{4.17 $\pm$ 0.04}     &   0.85$\pm$0.35   &   0.719  &  2.48  & 2.85 \\ \hline
Ours  & 1.89kbps   & 21.5 & \textbf{4.16 $\pm$ 0.04}    &     0.76$\pm$0.28   &   \textbf{0.738}    &  2.47 &  1.14 \\  \hline
Ours  & 1.78kbps   & 12.5 &  4.01 $\pm$ 0.04  &    \textbf{0.49$\pm$0.03}  &  0.635 & 1.06 & 1.01 \\ \hline
Ours  &  1.1kbps    & 12.5 &   4.02 $\pm$ 0.04      &  0.69$\pm$0.12   & 0.565  & \textbf{1}   &  \textbf{1}\\ \hline\hline 
% Ours  $+$ BPE &  1.78kbps    & 12.5 &   4.421  $\pm$     & 0.64  $\pm$  0.02   &  0.638 &1.06 & 1.01 \\ \hline
\begin{tabular}[c]{@{}l@{}} Ours - 10s \\ context \end{tabular} &  1.78kbps    & 12.5 &     3.84 $\pm$ 0.04   &   0.55 $\pm$ 0.07 &   0.691&   1.06 & 1.01 \\ \bottomrule

\multicolumn{6}{c}{\footnotesize * Note that  RTF and TTFA values are normalized  by setting the smallest value to 1.}
\end{tabular}
}
\vspace{-0.35cm}
\end{table}

Table \ref{tab:results-tts} presents the results of our ZS-TTS experiments. Koel-TTS trained with the 21.5 FPS NanoCodec achieved CER and MOS values comparable to those of the model trained with 21.5 FPS LFSC, while exhibiting better speaker similarity, a similar real-time factor, and a latency (TTFA) approximately 2.5 times shorter. In contrast, the model trained with the 12.5 FPS 1.78 kbps NanoCodec showed better intelligibility than its 21.5 FPS counterpart, while improving the real-time factor by 2.33 times and slightly reducing TTFA. However, it struggled to generalize to unseen speakers, leading to lower speaker similarity, and reduced overall quality, despite improving intelligibility. A similar trend was observed in the Koel-TTS model trained with the 12.5 FPS 1.1 kbps NanoCodec, which exhibited even lower speaker similarity and a slightly higher CER. This degradation is likely due to the reduced reconstruction quality of the 1.1 kbps NanoCodec compared to its 1.78 kbps variant. 

We hypothesize that the performance gap in speaker similarity and overall quality between the 12.5 FPS and 21.5 FPS models arises because the 12.5 FPS model generates nearly twice as much speech per autoregressive step, making the task more challenging. Conversely, the lower CER observed with the 12.5 FPS model could be attributed to the phoneme input and codec tokens having a similar rate, making alignment learning easier. Additionally, the shorter sequence length reduces the available speaker context by a factor of 1.7, potentially impairing speaker similarity. To mitigate this issue, we extended the context audio length from 5 to 10 seconds through repetition, which consistently improved speaker similarity from 0.635 to 0.691, but at the cost of a slight degradation in overall quality. These findings suggest that fully harnessing the potential of the 12.5 FPS NanoCodec may require architectural refinements to Koel-TTS.

\section{Conclusions and future work} \label{sec:conc}

In this work, we introduced NanoCodec, a state-of-the-art audio codec that achieves high-quality compression at a bitrate of 1.78 kbps and a frame rate of 12.5 frames per second. Experimental results demonstrate that NanoCodec surpasses existing approaches at the same bitrate, offering significantly higher intelligibility and speaker similarity. Furthermore, we showed that a Speech LLM trained with NanoCodec achieves quality comparable to that of a model trained with the LFSC, while exhibiting a lower real-time factor and reduced speech output latency, benefiting from our causal decoder. % Our ablation studies further reveal that bitrate reduction does not always improve downstream performance due to the trade-off between quality and intelligibility, emphasizing the need for careful optimization of compression efficiency and perceptual quality in Speech LLM applications.  
Our ablation studies further reveal that bitrate reduction does not always improve downstream performance due to the trade-off between quality and intelligibility. In future work, we plan to further investigate architectural modifications to Koel-TTS to fully leverage the potential of the 12.5 FPS NanoCodec.

\vfill\pagebreak
% \clearpage
\bibliographystyle{RefStyle}

\bibliography{references}

\end{document}